\documentstyle[11pt,newpasp,twoside,epsf]{article}
\def\edcomment#1{\iffalse\marginpar{\raggedright\sl#1\/}\else\relax\fi}
\reversemarginpar

\begin{document}
\title{Groups and clusters detection on DPOSS catalogs}
\author{E. PUDDU, V. STRAZZULLO, S. ANDREON, G. LONGO}
\affil{Osservatorio di Capodimonte, Via Moiariello 16, Napoli, ITALY}
\author{E. DE FILIPPIS}
\affil{Astrophysics Research Institute, Twelve Quays House, Egerton Wharf,
CH41 1LD Birkenhead, Wirral UK}
\author{R. GAL, S. DJORGOVSKI}
\affil{Dept. of Astronomy, Caltech, USA}
\author{R. SCARAMELLA}
\affil{Osservatorio di Monte Porzio, Roma, ITALY}

\begin{abstract}
We discuss the implementation and validation of a procedure aimed to detect 
groups and clusters of galaxies in DPOSS catalogs.
\end{abstract}

\section{Introduction}

The Digitised Palomar Sky Survey (hereafter DPOSS) offers an unique data
set to explore the large scale structure of the nearby ($z<0.2$) universe due
to its virtually unlimited sky coverage, to its deepness ($21.5$ in $g$) and
the availability of three colours (J, F and N).\\
In this poster we discuss a first application, on a small part of the whole 
data set, of a newly implemented procedure 
to search for candidate galaxy clusters and groups. Final goal of this 
project is a robust measurement of
the multiplicity function (hereafter MF) of galaxies.\\
The MF is a powerful tool to test the various cosmological scenarios. For
istance, if the protostructures form from a hierarchical gravitational
instability caused by primordial density fluctuations:
\begin{displaymath}  
\frac{d \rho}{\rho_{m}} \propto A \cdot M^{(-1/2 + n/6)}
\end{displaymath}
where $n$ is the initial spectrum index at era of baryonic recombination and
if one identifies groups of galaxies with mean density $r_{g}>k$ times the
average cosmological density $\rho_{M}$, then the function of the mass
distribution doesn't depend on the cosmic density parameter $\Omega$ but
only on the spectrum index $n$ (Gott \& Turner 1977). The groups which are seen now at a density 
enhancement
$k$ would therefore correspond to density fluctuations of amplitude $A$ at
recombination. Therefore, a measurement of the MF at the present epoch
derived from a catalogue of groups of galaxies identified for a given $k$,
gives a measure of the initial spectrum index $n$.

\section{Detection of candidate groups}

Even though on rather arbitrary ground, we define as a group all galaxy
ensembles having less then 15 members having total magnitude within 3
magnitudes from that of the brightest members. The implemented procedure
is described in detail in De Filippis (1999).\\
In order to compile a catalogue of candidate groups of galaxies in
absence of redshift information, we implemented a modified
version of the van Albada algorithm (Soares D.S.L. 1989). This algorithm makes
use of apparent magnitudes and projected
positions on the sky only, and gives for each pair of adjacent galaxies
their probability of being physically bound.\\
Assuming a Poisson statistic, the probability that the distance of a fixed
galaxy to its nearest non-physical companion lies between $\Theta$ and
$\Theta + d \Theta$, is given by:
\begin{displaymath}
P_{I}(\Theta) d \Theta = exp[- \pi \Theta^{2} n] 2 \pi \Theta n d \Theta
\end{displaymath}
The introduction of an adimensional distance x allows
to combine the angular separation of different pairs to a single
distribution removing the effects of clustering in the background.

\section{Detection of candidate clusters}

Starting points are the individual J, F and N catalogues produced by SKICAT
for a given DPOSS field after conversion to the g, r and i Gunn-Thuan
system. The three catalogues are then matched (assuming a maximum
matching distance of 7\arcsec, for details see Puddu et al. 1999).\\
After matching, all objects having $r>19.5$ are filtered out and the
resulting matched catalogue is binned into equal area square bins of
1.2\arcmin  to produce a density map. Then, S-Extractor (Bertin \& Arnouts
1996) is run in order to identify and extract all the
overdensities having number density $2 \sigma$ above the mean background
and covering a minimum detecting area of 4 pixs (4.8\arcmin  sq.). In this
way, as in the Schectman (1985) approach, we are not assuming any a priori
cluster model.
An application to the DPOSS field 610 is shown in Figure 1.\\
All the previously known Abell and Zwicky clusters are recovered and many
new candidates detected. For each candidate cluster we
then measure the S/N detection ratio and the Abell richness class.

\begin{figure}
\plotfiddle{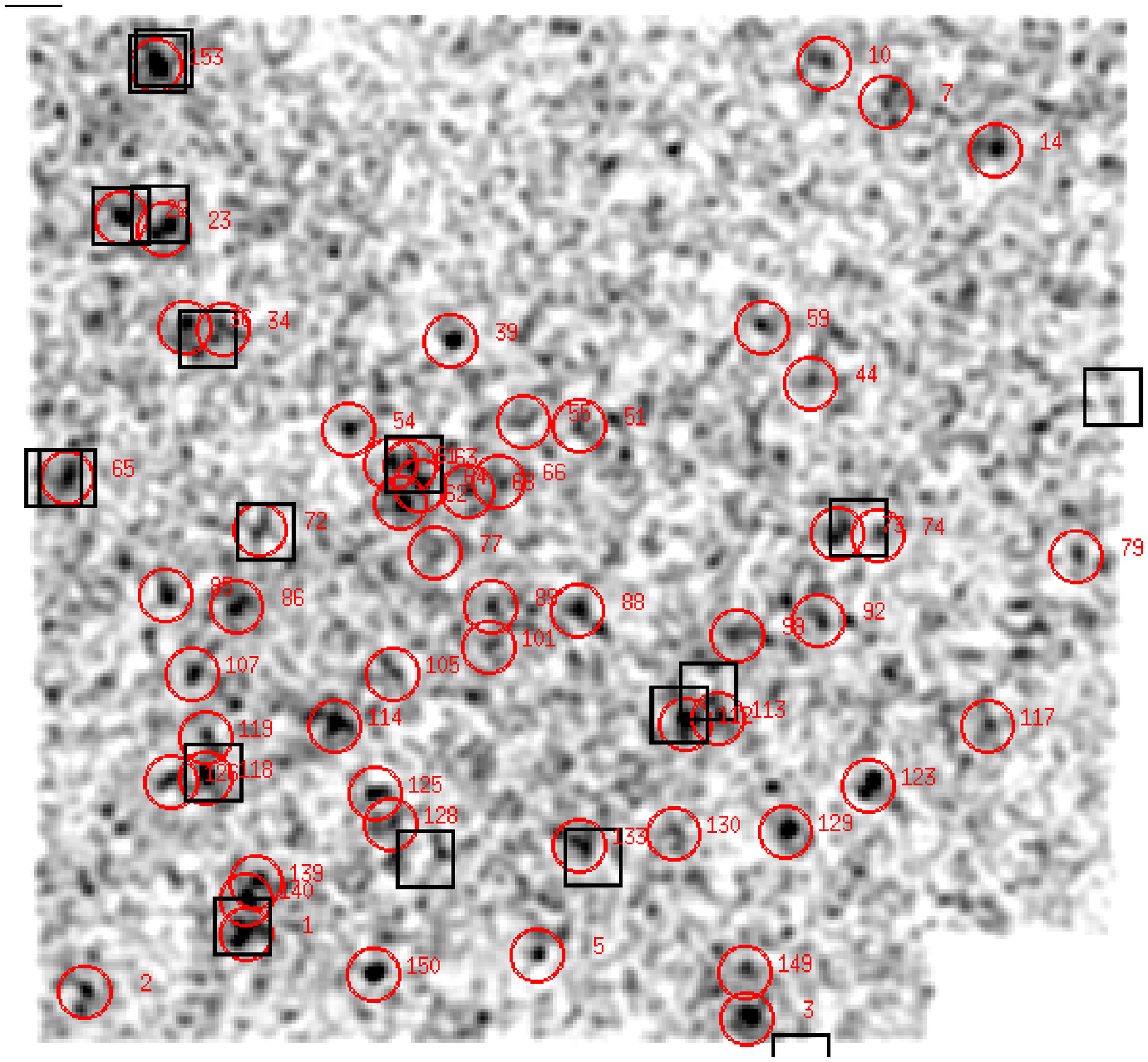}{8cm}{0}{50}{50}{-150}{-90}
\caption{Results for DPOSS field n. 610 (1\fh,+15\deg). Circles
mark the
position of detected candidate clusters while squares mark already
known Abell and Zwicky clusters.}
\end{figure}

\section{Validation of the algorithms}

In absence of a suitably complete samples of galaxies with known redshift
we were forced to validate the algorithms shortly summarised above using
either simulations or photometric techniques. In what follows we shall
discuss first the groups and then the clusters.

\subsection{Group algorithm validation}
In order to test the accuracy (lost groups) and reliability (spurious
groups) of the algorithm we tested it on 70 realistically simulated
fields. We first produced the galaxy background by assuming uniform
distribution and the field luminosity function given by Metcalfe et al.
1995; then we added groups of galaxies according to the multiplicity
function by Turner E.L. \& Gott J.R. 1976, with redshift computed
according to: 
\begin{displaymath}
N(z)=\frac{32 \pi\rho_{0} c^{3}}{3 H_{0}} \left[\frac{1}{z+1}
\left(1 - \frac{1}{\sqrt{(1+z)}}\right)\right]^{3}
\end{displaymath}
and absolute
magnitude of the brightest galaxy in the group taken from the cumulative
LF for groups of galaxies: 
\begin{displaymath}
\Phi(M) dM = \Phi^{*} [10^{0.4(M^{*}-M)}]^{(\alpha + 1)} \cdot exp[-10^{
0.4(M^{*}-M)}] dM
\end{displaymath}
where $M^{*} =-20.85$ and $\alpha=-0.83-+0.17$. 
Other parameters varied in the course of the
simulations were: i) the diameter of the group inside a gaussian
distribution centered at $D_{0}=0.26Mpc$; ii) the maximum possible redshift
for a group (chosen in the range $0.2< z <0.7$). Different simulations
have been performed changing the lower limit of the probability for wich
two galaxies are considered physical companions by the algorithm. The best
compromise between accuracy and reliability is found at $p=0.6$.\\
In Figure 2 we show the multiplicity function obtained from 12 DPOSS
plates (covering a total solid angle of $483 deg^2$) compared to six
other MFs taken from the literature:
MF derived from magnitude limited surveys making use of redshift
information,
the CfA survey (Geller \& Huchra 1983; Garcia et al. 1993;
Ramella \& Pisani 1997; Ramella et al. 1998),
MF derived from diameter limited survey (Maia \& Da Costa 1989) and
MF derived from limited magnitude survey without redshift informations
(Turner \& Gott 1976).\\
As it can be clearly seen, the MF derived by appling our method to DPOSS
material is at $98\%$ confidence level undistinguishable from literature
MF obtained from redshift surveys.

\begin{figure}
\plotfiddle{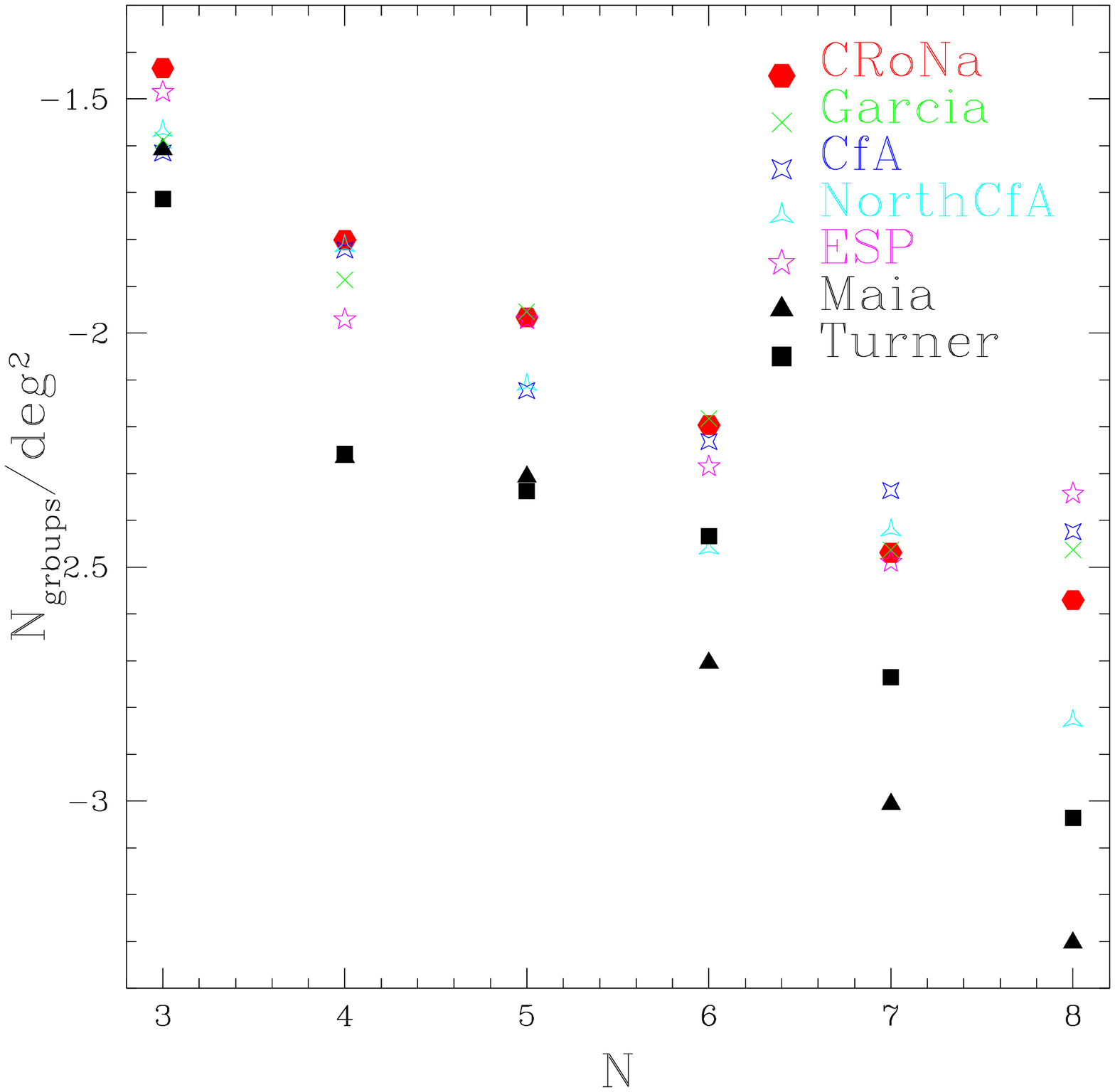}{8cm}{0}{40}{40}{-150}{-60}
\caption{Comparison of our MF against those derived by other authors.}
\end{figure}

\subsection{Cluster algorithm validation}

We took two different approaches: through simulation and through magnitude
- color diagrams. In order to produce realistic simulations we 
followed a rather lengthy procedure. First of all, we needed a fairly
accurate multiplicity function to be used as input parameter for the
simulation. Since there is no good MF available in literature, we used a
set of 20 DPOSS triplets to identify clusters and derive a preliminary MF
(we used only very high S/N clusters for which there is no doubt about
their physical nature).\\
In order to simulate the fields we first distribute randomly the cluster
centers and we assume the cluster distribution to
be uniform in the volume element and randomly assign to each cluster a
redshift in the range $[0.02,0.2]$.
Then we grow the clusters around their centers randomly selecting a radius
in the range $[2-3]$ Mpc and weighting the number of members accordingly to
the above derived MF.
Absolute magnitudes for galaxies are then extracted from a
Schechter function ($M^{*} =-21.41,\alpha = -1.24$) and scaled to apparent
magnitudes taking into account the K-correction term.
We finally add field galaxies using the Metcalfe field counts and LF.
Even though simulations are still in progress, preliminary results show
that our algorithm can recover $95\%$ of the clusters at $z<0.15$ and
$70\%$ of the clusters in the range  $[0.15,0.2]$.\\

\subsection{Photometric validation}

As stressed above, DPOSS data are calibrated via independent CCD frames
taken at various telescopes in the $1-2m$ range. This means that we have at
our disposal a large number of deep fields, where galaxy photometric
properties can be measured with high photometric accuracy. In the last two
years we selected calibration fields in order to largely overlap with our
candidate clusters sample.\\
These data were used to derive color - magnitude diagrams.\\
To calibrate the CCD frames taken in the $g$, $r$ and $i$ Gunn--Thuan system, we
use S-Extractor to identify galaxies and derive the colour-magnitude
diagram ($r$ vs. $g-r$) for the candidate cluster region and for a test field
(background) well selected. Then we statistically subtract the field
contamination and identify the excess of galaxies as cluster members. 
The existence of an excess of galaxies in the putative cluster line of
sight and the fact that they obey to the usual colour--magnitude relation 
is taken as confirmation of the existence of the putative cluster.
These galaxies are then used
to build the cluster radial profile. The colour-magnitude diagram allows
also to derive a rough photometric redshift (which, due to the broad band,
has however a rather large error).\\
Preliminary results confirm that for $z<0.15$ the candidate clusters
found by our algorithm are $95\%$ "true" clusters. At higher redshift
($0.15<z<0.2$) this fraction drops to $70\%$.

\begin{figure}
\plottwo{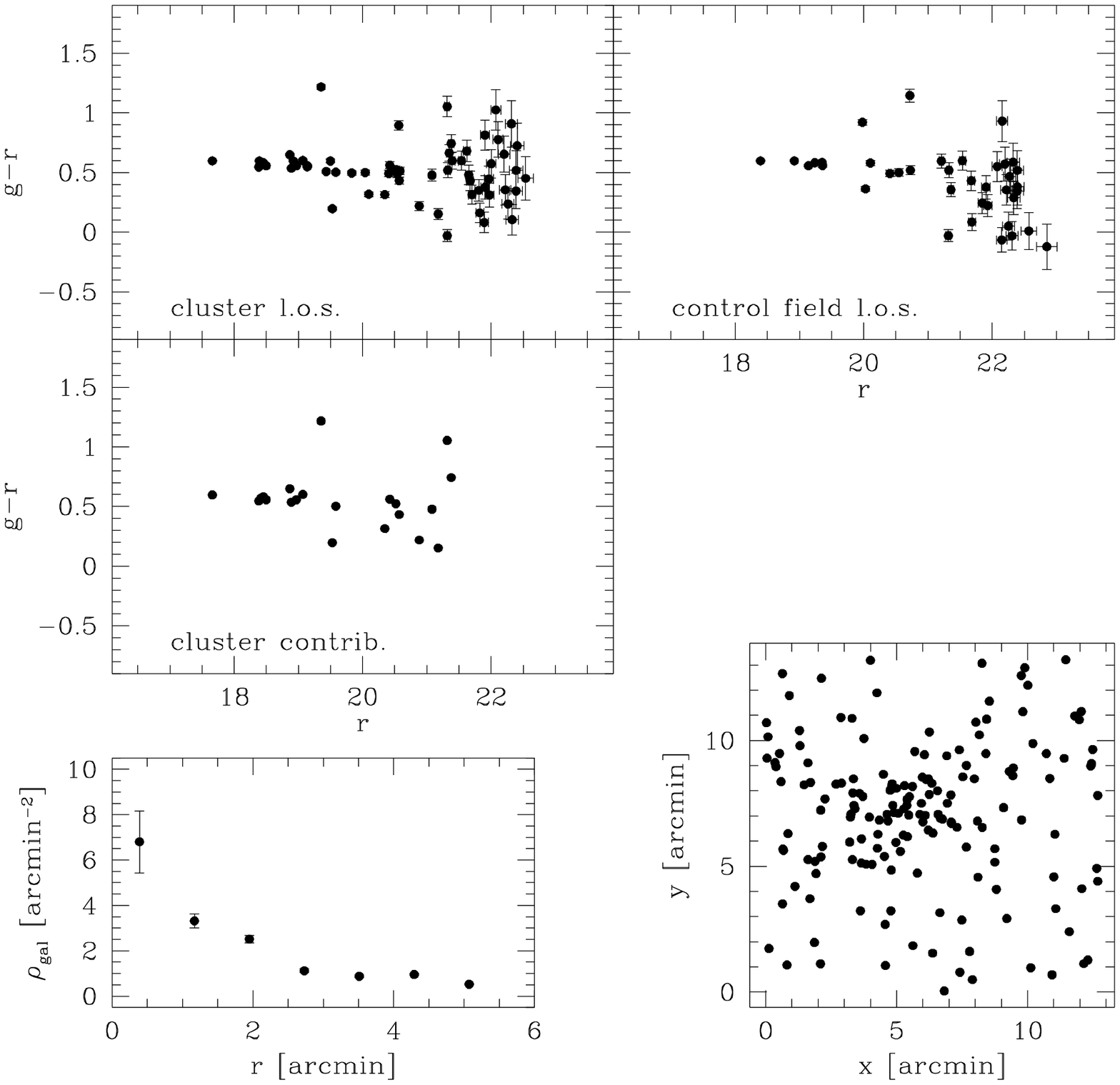}{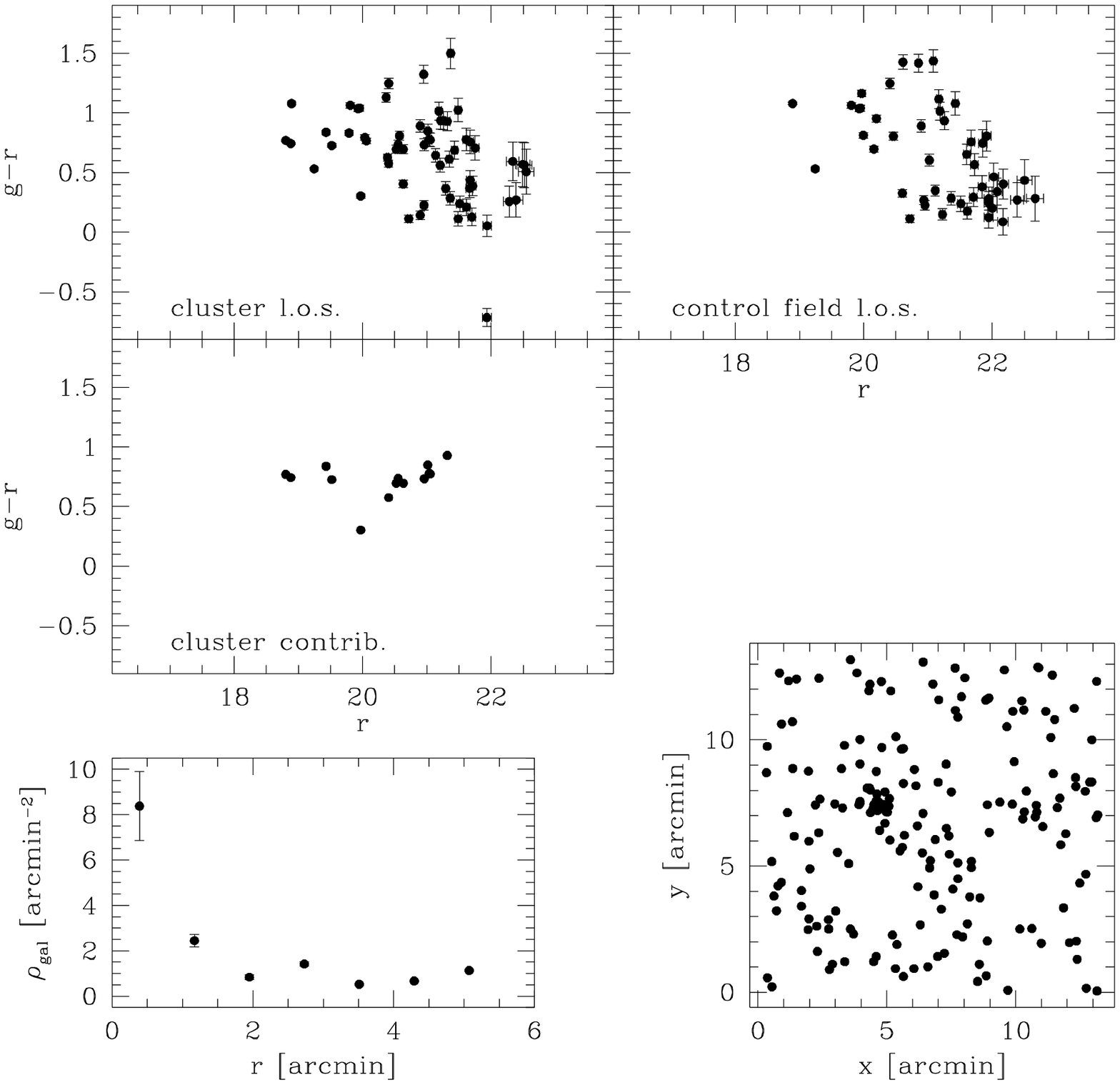}
\caption{Color - magnitude diagram for the
candidate cluster obtained accordingly to the procedure described in the
text. Right (high S/N candidate): the cluster turns out to be of richness class I 
and to be at $z=0.10-0.12$. Left (low S/N candidate): the cluster turns out to be of richness class I 
and to be at $z=0.26-0.3$.}
\end{figure}

We plan to improve the simulation by adding non-circular simmetry for the
simulated cluster and to study the dependence on the zero-order
correlation functions. We also plan to apply the colour-magnitude diagram
test to all clusters avalaible in our calibration data set.

\end{document}